# Magnetism and anomalous transport in the Weyl semimetal PrAlGe: Possible route to axial gauge fields

Daniel Destraz[1*], Lakshmi Das[1], Stepan S. Tsirkin[1], Yang Xu[1], Titus Neupert[1], J. Chang[1*], A. Schilling[1], Adolfo G. Grushin[2], Joachim Kohlbrecher[3], Lukas Keller[3], Pascal Puphal[4], Ekaterina Pomjakushina[4] and Jonathan S. White[3*]

[1]*Physik-Institut, Universität Zürich, Winterthurerstrasse 190, CH-8057 Zürich, Switzerland*

[2]*Univ. Grenoble Alpes, CNRS, Grenoble INP, Institut Neel, 38000, Grenoble, France*

[3]*Laboratory for Neutron Scattering and Imaging (LNS), Paul Scherrer Institute (PSI), CH-5232 Villigen, Switzerland*

[4]*Laboratory for Multiscale Materials Experiments (LMX), Paul Scherrer Institut (PSI), CH-5232 Villigen PSI, Switzerland*

\* To whom correspondence should be addressed. E-mail: destraz@physik.uzh.ch, johan.chang@physik.uzh.ch, jonathan.white@psi.ch

**Abstract**

In magnetic Weyl semimetals, where magnetism breaks time-reversal symmetry, large magnetically sensitive anomalous transport responses are anticipated that could be useful for topological spintronics. The identification of new magnetic Weyl semimetals is therefore in high demand, particularly since in these systems Weyl node configurations may be easily modified using magnetic fields. Here we explore experimentally the magnetic semimetal PrAlGe, and unveil a direct correspondence between easy-axis Pr ferromagnetism and anomalous Hall and Nernst effects. With sizes of both the anomalous Hall conductivity and Nernst effect in good quantitative agreement with first principles calculations, we identify PrAlGe as a system where magnetic fields can connect directly to Weyl nodes via the Pr magnetization. Furthermore, we find the predominantly easy-axis ferromagnetic ground state co-exists with a low density of nanoscale textured magnetic domain walls. We describe how such nanoscale magnetic textures could serve as a local platform for tunable axial gauge fields of Weyl fermions.

**Introduction**

In Dirac and Weyl semimetals, the emergence of large Berry curvatures due to electron band degeneracies, or singular band touching points (Weyl nodes), leads to striking topological effects on conduction electrons[1,2] that can be detected by measurements of negative magnetoresistance[3], and the anomalous Hall effect (AHE)[4,5]. In magnetic semimetals, it is known that both the size and direction of the magnetization can generate Weyl nodes and shift their positions, which provides the possibility for magnetic field control of the Weyl node positions and the emergence of the AHE, a critical aspect of the nascent field of topological electronics[6].

Experimentally, recent studies connecting the size of the AHE to Weyl-node induced Berry curvature in magnetic systems include antiferromagnetic (AFM) $Mn_3Sn$ [7,8], kagome ferromagnetic (FM) semimetal $Co_3Sn_2S_2$ [5], and the topological nodal line ferromagnetic semimetal $Fe_3GeTe_2$ [9]. Similarly, other works have connected the anomalous Nernst effect to the Berry curvature[10,11]. In these studies, a theoretical description of the observations provides the direct link between topological properties of the band structure and magnetically sensitive observables. This motivates both experiments and computational material science aimed at the rational engineering of band structures and the prediction of new systems with magnetically sensitive topological properties.

The rare-earth compounds $R$AlGe ($R$ = Ce, Pr, La) with polar tetragonal $I4_1md$ crystal structure were proposed recently to host either type-I and/or type-II Weyl fermions depending on the rare-earth ion[12]. With CeAlGe [13] and LaAlGe [14] respectively proposed to be magnetic and nonmagnetic type-II Weyl semimetals, PrAlGe is expected to be a magnetic type-I Weyl semimetal with space-inversion symmetry broken by the polar lattice symmetry, and time-reversal symmetry broken at the onset of an easy-axis ferromagnetic ordering along *c* [Fig. 1a][12]. As illustrated schematically in Figs. 1b and c, the locations in momentum space of Weyl nodes are expected to vary with the size of the magnetization into a time-reversal symmetry breaking configuration. An observable consequence is the emergence of anomalous Hall conductivity (AHC) in the plane normal to the direction of the magnetization, which is due to the associated Berry curvature that remains finite when evaluated over the Brillouin zone below the Fermi energy.

Here we apply a range of experimental techniques combined with first principle calculations to determine the relationship between magnetism and anomalous transport in PrAlGe summarized in Fig. 1d. Neutron scattering experiments reveal that below the critical temperature $T_c$, PrAlGe orders predominantly as an easy-*c*-axis ferromagnet. Notably we obtain further evidence that in part of the sample Pr moments tilt away from *c* over a characteristic nanometric length scale, which we interpret to be due to large FM domain walls. From electrical transport, we find the ground state displays no spontaneous anomalous Hall or Nernst signals in zero magnetic field. Both responses emerge sharply as small magnetic fields polarize the Pr moments along *c*. As seen in Fig. 1d, the magnetic field derivative of the Hall resistivity $d\rho_{xy}/dB$ is large and positive for $T < T_c$ and magnetic fields $B < B_s$, where $B_s$ is the saturation field. This observation signifies a contribution to the Hall resistivity beyond the usual Hall effect due to the anomalous Hall effect (AHE). We discuss the origin of the observed AHE and furthermore an anomalous Nernst effect (ANE), and provide theoretical support that the origin of both is due to the Weyl node-induced Berry curvature. PrAlGe is thus a magnetic Weyl semimetal that displays easily controllable transverse electronic responses using low applied magnetic fields that couple to the Pr magnetization.

**Results**

**Neutron scattering**

To determine the magnetic order in PrAlGe, we obtained powder neutron diffraction (PND) data at DMC, Paul Scherrer Institute (PSI) shown in Fig. 2a. Upon cooling below $T_c$, extra scattered intensity due to magnetism appears at scattering angles commensurate with the tetragonal point symmetry of the PrAlGe lattice, and can be described by a propagation vector $\mathbf{Q} = 0$. Within a standard magnetic symmetry analysis, the $\mathbf{Q} = 0$ propagation vector and spacegroup symmetry lead to three symmetry-allowed magnetic structure models that can be tested against the data. Fig. 2b shows a successful refinement of the PND profile obtained at 1.6 K, which includes both the nuclear and magnetic scattering. The model for magnetic order is the standard irreducible representation $\Gamma^1$ that describes a ferromagnetic (FM) order with moments aligned with the $c$-axis [Fig. 1a]. Magnetic structure refinements according to the other allowed irreducible representations do not describe the observed magnetic scattering and are disregarded. This result shows that below $T_c$ PrAlGe displays ferromagnetic correlations along the easy $c$-axis. The size of the ferromagnetic moment refined from the data is 2.29(3) $\mu_B$/Pr.

To investigate the magnetic correlations further, small-angle neutron scattering (SANS) measurements were performed on a PrAlGe single crystal using the SANS-I instrument, PSI. SANS probes variations in the magnetization density on the meso- to nanoscale, covering a typical range of real-space length scales from 5 to 500 nm. Therefore, the observation of magnetic SANS at low momentum transfers near $|q| = 0$ (which is the FM zone center) can signify the existence of short-range spin correlations and/or magnetic spins correlated over nanometric length scales. From PrAlGe we indeed observe significant magnetic SANS to emerge for $T < T_c$ and for $B < B_s$ as shown in Figs. 2c-g. This observation suggests that in addition to dominant easy-$c$-axis ferromagnetic correlations, a co-existing nanoscale magnetic texture also exists in the sample below $T_c$.

Fig. 2c shows the typical SANS pattern obtained at 1.9 K and $\mu_0 H = 0$. The intensity is distributed uniformly in azimuth around the origin, and falls monotonically over an extended range of $|q|$. Fig. 2e shows the azimuth-averaged one-dimensional profiles of SANS intensity versus $|q|$ at various temperatures during a $T$-warming scan. Profiles obtained at T > $T_c$ overlap and show no $T$-dependence, thus representing background scattering. Subtracting the high-$T$ background from the data obtained for $T < T_c$ leaves the magnetic scattering, the profile integral of which is shown at each $T$ in Fig. 2f. The extracted intensity clearly disappears at $T_c$ confirming its magnetic origin. A power law fit of the data near to $T_c$ according to $I \propto (1-T/T_c)^{2\beta}$ yields $T_c$ = 15.1(1) K and $\beta$ = 0.30(1). The latter is reasonably close to the value of 0.325 expected for the 3D Ising model[17]. Finally, Figs. 2d and 2g show SANS data obtained after applying a magnetic field along the easy $c$-axis ($\mu_0 H \parallel c$) after a ZFC to 1.9 K. The magnetic SANS intensity is largely suppressed by ~0.3 T as the system is driven field-polarized.

**Thermodynamic and transport measurements.**

Zero-field resistivity data shown in Fig. 3a reveal the ground state of PrAlGe is metallic. The flux-grown single crystal displays a residual resistivity of $\rho_0$ = 102 μΩ cm and a modest residual resistivity ratio $\rho(300\ K)/\rho_0 \sim 2$. These values are comparable with those determined previously from resistivity measurements on a floating-zone grown sample[18], where it is found that $\rho_0 = 228$ μΩ cm and $\rho(300\text{ K})/\rho_0 \sim 1.7$. Below $T_c$, Fig. 3a shows that the resistivity scales as $T^3$. This scaling has a remarkably sharp cut-off at $T_c$ and above this temperature the resistivity is well described by the standard Bloch-Grüneisen formula (minus a $T^3$ correction term).

Fig. 3b shows the dc magnetization obtained on a floating-zone grown PrAlGe crystal for magnetic fields applied along the *c*-axis. At low fields, we observe a steep rise of the magnetization upon cooling below $T_c$, followed by a further rise around 12 K suggested previously to be due to a spin-reorientation transition[18], and finally a weaker *T*-dependent plateau down to base *T*. With increasing field strength, the sharp features broaden, and the plateau saturates at 2.3 $\mu_B$/Pr.

For fields above 0.1 T until saturation $B_s \sim 0.35$ T, the magnetization data display a characteristic temperature scale $T_g \sim 10$ K, below which the magnetization is history dependent and field-cooled and zero-field-cooled (FC and ZFC) curves bifurcate. This feature is consistent with $T_g$ corresponding to a spin-freezing transition temperature. In Supplementary Note 1, we characterize the regime below $T_g$ further, presenting evidence in Supplementary Fig. 1 for the so-called thermoremnant magnetization effect characteristic of slow spin glass-like dynamics[19,20].

For fields of 0.1 T and higher, Fig. 3b shows $T_g < T_c$, consistent with reentrant spin glass behavior[19,20]. Previous work shows that at fields below 0.1 T $T_g$ increases, and tends towards $T_c$ in mT fields[18]. In the reentrant spin-glass picture, the magnetic ground state displays a co-existence of spin components described as longitudinal and transverse with respect to an applied field. Here the longitudinal spin component is aligned with the Ising anisotropy axis, and leads to both the bulk magnetization and the ferromagnetic Bragg peaks observed by PND. The spin-glass properties are expected to arise from a freezing at $T_g$ of transverse spin components tilted randomly from the longitudinal axis. In principle, the tilted components can display short-range magnetic correlations observable as diffuse magnetic neutron scattering near to the ferromagnetic Bragg peaks[21,22]. From PrAlGe, we find agreement between the size of the refined longitudinal Pr moment from the PND data and that determined from the bulk magnetization, and there is no clear signature of diffuse scattering in the range of momentum transfer explored by PND. Therefore, we conclude that the transverse spin components are small enough such that any associated correlations lie below our detection limit. In turn, this observation supports the allocation for the origin of the magnetic SANS signal at low momentum transfer as less likely due to short-range correlations, and instead more likely due to a nanoscale magnetic texture, as we discuss later.

In Fig. 3c we show the thermal dependence of the Hall resistivity $\rho_{xy}$ from a flux-grown crystal at various magnetic fields along the *c*-axis to display a strong resemblance to the magnetization curves. In the low-field regime, both Hall resistivity and magnetization display a sharp upturn across $T_c$ and history-dependence below $T_g$ - see also Supplementary Fig. 1. This, combined with the observation of a pronounced kink in the resistivity at $T_c$, confirms the direct coupling between the conduction electrons and the microscopic magnetism.

**Anomalous Hall and Nernst effect in PrAlGe.**

Next we explore the coupling between itinerant and localized electrons from both Hall and Nernst effect measurements. Above $T_c$, both Hall and Nernst isotherms vary linearly for magnetic fields applied along the $c$-axis [Figs. 4a and 4b)]. The Hall coefficient $R_H = \rho_{xy}/\mu_0 H$ is positive and increases mildly on cooling from 100 K down to $T_c$. Since $R_H > 0$ hole-like carriers most likely dominate the charge transport. While the observed Hall coefficient is likely the result of compensated electron- and hole-like transport, it is interesting to apply a single band model $R_H = 1/ep$ where $p$ is the hole carrier concentration. At $T = 100$ K, we obtain $p = 7.8 \times 10^{20}$ cm$^{-3}$ ($E_F = 310$ meV), placing PrAlGe in the semi-metallic regime. In a similar fashion, we obtain the Nernst coefficient $\nu = N/B = 20$ nV / (K T) at $T = 21$ K, where the Nernst signal $N$ is the ratio of the transverse electrical field and the longitudinal temperature gradient $N = -E_y / (\nabla_x T)$. With a Hall mobility $\mu = 8 \times 10^{-3}$ T$^{-1}$, this corresponds to $\nu/T = (4.6 \times 10^{-4}$ V K$^{-1}) \mu/T_F$, falling close to the universality curve [$\nu/T = (2.83 \times 10^{-4}$ V K$^-$ [1]) $\mu/T_F$] for the quasiparticle Nernst effect[23,24]. Above $T_c$, PrAlGe thus behaves as a standard `dirty' semimetal.

Below $T_c$, both the Hall and Nernst isotherms display a non-linear dependence on magnetic field. Compared with the regime above $T_c$, both the Hall and Nernst coefficients display a five-fold increase in the $\mu_0 H \to 0$ limit. In the field-polarized regime, the slope of the Hall resistivity and Nernst signal each fall back to values comparable to those found high above $T_c$, where no influence of the anomalous contribution persists. For the Hall effect this is also illustrated by plotting $d\rho_{xy}/dB$ in Fig. 1d. Whereas we observe no spontaneous Hall and Nernst signals in zero magnetic field, their non-linearity with magnetic field and their correlation to the magnetic properties strongly suggest magnetism to drive the emergence of anomalous transport signals.

To quantify the sizes of the observed anomalous Hall and Nernst effects in zero field, one way is by extrapolation from above the saturation field to the $\mu_0 H \to 0$ limit. In this fashion, an anomalous Hall conductivity $-\sigma^A_{xy} = \rho_{xy}/(\rho^2_{xy} + \rho^2_{xx}) \approx \rho_{xy}/\rho^2_{xx} = 367\ \Omega^{-1}\ cm^{-1}$ is found at 2 K. An extrapolated anomalous Nernst response $N^A \approx 28$ nV / K is found at $T = 11$ K. As shown in Fig. 4c, both the onset of anomalous Hall and Nernst effects coincide with the onset of magnetism. The solid line corresponds to a power law fit of the anomalous Hall conductivity to the same equation as used for fitting the $T$-dependent SANS intensity [Fig.

2c]. The fitted exponent $\beta = 0.28(4)$ is in good agreement with the value obtained from the SANS data, reinforcing the conclusion that the anomalous transport is sensitive to local moments. In Supplementary Note 3 and Supplementary Fig. 3, we compare the sizes of $\sigma^A_{xy}$ and $N^A$ in PrAlGe with those observed in other semimetals, and find them to lie within the ranges typical of known relevant systems.

For comparison with theory, it is useful to connect the anomalous Nernst response $N^A$ with the off-diagonal Peltier coefficient $\alpha_{xy}^A = \sigma_{xx} N^A + \sigma_{xx} S(\sigma_{xy}^A/\sigma_{xx} + \kappa_{xy}^A/\kappa_{xx})$[25,26]. Generally, insights into $\alpha_{xy}^A$ require measurements of the Seebeck coefficient $S$ and thermal conductivity $\kappa$. In the present case of PrAlGe, Supplementary Fig. 2 shows $S \approx 0$ at $T \approx 14$ K. Thus in this temperature range, it is reasonable to assume $\alpha_{xy}^A \approx \sigma_{xx} N^A$. At $T = 14$ K, we therefore estimate $\alpha_{xy}^A/T \approx 14$ μV / (K$^2$ Ω cm). We note that due to the poor mobility, the normal state quasiparticle Nernst response is at a level where this relatively modest anomalous Nernst signal is accessible in our experiment.

**Discussion**

Based on our experimental results, we find that below $T_c$ PrAlGe displays easy-$c$-axis ferromagnetism, with a refined ferromagnetic moment close to that measured from the bulk magnetization. From SANS in particular, we evidence that a low volume of regions (estimated to be less than 1%) where the magnetization tilts away from $c$ over a nanometric length scale that enriches the magnetism. A minimal model for the $|q|$-dependence of the SANS intensity is one that describes a cycloidal-like tilting of neighboring moments away from the $c$-axis and toward in-plane directions[27,28] (see Supplementary Note 4 and Supplementary Figs. 4-6 for more details). Physically this description is compatible with being due to nanoscale ferromagnetic domain walls. Fig. 2e shows a global fit of the model to all SANS profiles that provides a successful description of both the $|q|$-dependence of the intensity and a quantification of the nanometric magnetic length scale that varies from ~14 nm at 2 K to ~25 nm close to $T_c$ (see Supplementary Fig.5).

In addition to a cycloidal-like tilting of moments, the SANS data at hand may also have contributions from the afore-mentioned short-ranged transverse spin correlations that form either within ferromagnetic domains or around crystallographic defects. Further studies by microscopy and polarized SANS techniques can lead to a more precise description of the nanoscale magnetization texture. Nonetheless, an important common feature among all candidate models is the existence of magnetization components tilted from the $c$-axis. While this could be considered surprising due to the clear easy-axis anisotropy of the system, non-collinear moment arrangements may arise due to either symmetry-allowed Dzyaloshinskii-Moriya interactions in the $I4_1md$ spacegroup, and/or competing interactions between in-plane

AFM or out-of-plane FM interactions[18]. The latter can also be responsible for the observed spin glass-like behaviour below $T_g$ similarly as for other Pr magnets[29-32].

Next we turn to the origin of anomalous transport, focusing firstly on the anomalous Hall effect (AHE). In general, an AHE arises in any FM in the presence of spin-orbit coupling and includes both intrinsic (due to Berry curvature) and extrinsic (due to scattering) contributions. In FM Weyl semimetals, the origin of the intrinsic AHE is particularly transparent: it originates from the Weyl node-induced Berry curvature monopoles[2]. Here we present first principles calculations to estimate the intrinsic $\sigma_{xy}^A$ in PrAlGe for a uniform magnetization along $c$, and the limit $T \to 0$. In accordance with Ref. 36, extrinsic effects are not expected for our observed magnitude of $\sigma_{xx}$. Other scenarios that can also lead to a large Berry curvature (and hence AHE) include a presence of massive Dirac fermions[33] or magnetic nodal lines[34,35]. Due to the symmetry of the present crystal structure however, such effects are not expected for PrAlGe.

Following Ref. 12, we calculated the band structure of PrAlGe finding multiple Weyl nodes existing close to the Fermi energy that do not lie on high symmetry points in the Brillouin zone (see Fig. 5a, Supplementary Note 5 and Supplementary Fig. 7 for more details). The anomalous hall conductivity is evaluated as

$$\sigma_{ab}^A = -\frac{e^2}{\hbar} \int [dk] \sum_n \theta(\mu - E_{kn}) \epsilon_{abc} \Omega_{kn}^c \qquad (1)$$

$$\Omega_{kn} = \nabla_k \times A_{kn} = -\text{Im}\langle\partial_k u_{kn}|\times|\partial_k u_{kn}\rangle \tag{2}$$

where $A_{kn} = i\langle u_{kn} \vee \partial_k u_{kn}\rangle$ is the Berry connection, $\Omega_{kn}$ is the Berry curvature, $E_{kn}$ is the band energy, $\theta$ is the Heaviside step function, and the integral is over the Brillouin zone with $[dk] \equiv d^3k / (2\pi)^3$. When the magnetization is along the $z$-axis, $\sigma_{xy}$ is determined by the component $\Omega^z$ of the Berry curvature. Figs. 5a and 5b show how the distribution of Weyl nodes affects the Berry curvature (and $\sigma_{xy}^A$), by plotting the calculated momentum space ($k$-)resolved $\Omega^z$ obtained by summing over states below the Fermi level. Fig. 5b shows the $k_z$-resolved Berry curvature that is integrated in the $xy$-plane. We see sharp steps near $k_z = 0$ and $k_z \sim 2\pi/c$, indicating the presence of Berry curvature dipoles. These dipoles are formed by pairs of Weyl nodes of opposite chirality, separated in momentum space along the $z$ direction. Note that if all the Weyl nodes would be exactly at the Fermi level, this curve would have sharp integer-valued steps at the $k_z$ coordinates of the Weyl points. However, since the Fermi surface has finite size, these steps are smooth. Fig. 5a shows a colormap of the Berry curvature in the $k_x$-$k_y$ plane together with the projection of the positions of the Weyl points. The large intensity around the Weyl nodes that are located close to the Fermi level demonstrate that they are indeed an important source for the observed anomalous Hall conductivity. Our calculations further reveal that large Berry curvature can also exist near avoided crossings that arise generally in the band structure, see Supplementary Note 5 and Supplementary Figure 7 for further details.

Fig. 5c shows the calculated $\sigma_{xy}^A$ as the chemical potential is varied over a range of ±0.1 eV around the calculated Fermi level. By varying the chemical potential, we take into account possible effects due to doping of the sample, or inaccuracies in the calculated band structure. We observe that the experimental estimate for $\sigma_{xy}^A$ at 2 K [red diamond in Fig. 5c] lies in broad agreement with those expected for the explored range of chemical potential, in particular for $\mu = E_F$, where the theoretical estimate is $\sigma_{xy}^A = 330$ $\Omega^{-1}$ cm$^{-1}$. The agreement between the experimental and the calculated values for $\sigma_{xy}^A$ supports the conclusion that the dominant origin for the observed anomalous Hall effect is the large Berry curvature induced by the Weyl nodes.

To consider a theoretical estimate of the Nernst coefficient, we computed $\alpha_{xy}$ as[37]

$$\alpha_{xy} = \frac{-1}{e} \int d\varepsilon \frac{\partial f}{\partial \mu} \sigma_{xy}(\varepsilon) \frac{\varepsilon - \mu}{T} \qquad (3)$$

which relates the Hall conductivity to the off-diagonal Peltier coefficient $\alpha_{xy}$ and is valid also for the anomalous components. One can see in Fig. 5c that in contrast to the AHC which depends weakly on the chemical potential, the coefficient $\alpha_{xy}$ fluctuates rapidly between positive and negative values. At $\mu = E_F$ we find $\alpha_{xy}^A/T \approx 20$ µV / (K$^2$ $\Omega$ cm), which is marked by a magenta diamond in Fig. 5c, and is a factor 1.4 greater than the experimental value. Taking into account the complexity of both experiment and numerical calculation, this can be considered as good level of agreement. However, one should bear in mind that the strong dependence of $\alpha_{xy}^A$ on the chemical potential suggests that this quantity is also very sensitive to tiny details of the band structure, which probably lie beyond the accuracy of any modern ab initio method. On the other hand, the order of magnitude estimate $|\alpha_{xy}^A/T| < 100\ \mu V / (K^2\ \Omega\ cm)$ may be considered as a reliable estimation.

Finally, we discuss an important implication of our results namely that in magnetic Weyl semimetals, spatially inhomogeneous magnetism such as that implied from the SANS data here in PrAlGe, may serve as a platform for tunable axial gauge fields of Weyl fermions. When coupled to electromagnetic gauge fields, Weyl semimetals display phenomena distinct to trivial metals due to the strong Berry field in momentum space near the Weyl nodes[2]. When considered as relativistic fermions, different external perturbations on a Weyl semimetal find a unified description in terms of additional gauge fields coupled to the Weyl fermions. In particular, inhomogeneities can act as an axial gauge field which, unlike the real electromagnetic field, distinguishes the node chirality[38].

The effects of axial magnetic fields have been realized in synthetic systems; two-dimensional arrays of CO molecules on copper[39], and three-dimensional acoustic metamaterials and photonic systems[40,41]. In hard condensed matter systems, however, axial fields have been observed in two-dimensional graphene sheets[42], but never in three-dimensional systems. The spatially varying magnetization observed by SANS here in PrAlGe potentially makes PrAlGe a natural host of tunable axial gauge fields at low temperatures and low magnetic fields[43]. Since a spatially varying magnetization profile in a Weyl semimetal is tantamount to a position-space-dependent Weyl node separation in momentum space, the variation of the Weyl node positions defines axial gauge fields in time-reversal breaking Weyl semimetals[43-45]. Our measurements indicate that the proliferation of spatial inhomogeneities (i.e., domain walls) is tunable by changing the temperature and the external magnetic field. This suggests that PrAlGe is not only a magnetic Weyl system, but also a platform to tune axial gauge fields in a solid-state system. As discussed further in Ref. 38, unique experimental signatures due to axial gauge fields are expected in quantum oscillation and transport measurements. Since domain averaging precludes the observation of the effects of axial gauge fields in the present bulk measurements, the future challenge is to probe the system on the local scale of a single domain, or equivalently a single nanometric domain wall, such that the signatures of axial gauge fields may be observed, and potentially exploited.

In summary, we applied neutron scattering, magnetization, resistive and thermoelectric transport techniques to elucidate the direct link between magnetism and the anomalous Hall and Nernst effects in the magnetic Weyl semimetal PrAlGe. From transport data obtained in an easy *c*-axis magnetic field, we determined the size of the anomalous Hall conductivity $-\sigma_{xy}^A = 367$ $\Omega^{-1}$ cm$^{-1}$ at 2 K, and the anomalous Nernst signal $N^A \approx 28$ nV / K at 11 K. The sizes of $-\sigma_{xy}^A$ and $N^A$ are in good agreement with those obtained from first-principle calculations of the electronic structure in the presence of uniform *c*-axis magnetization, and allows us to allocate their origin as due to the Berry curvature distribution. From SANS measurements in particular, we obtain further evidence in the ferromagnetic ground state for a small fraction of a co-existing nanoscale-sized magnetization texture involving moments tilted away from *c*, which is likely due to nanoscale ferromagnetic domain walls. The existence of such nanoscale domain walls in a Weyl semimetal could make PrAlGe a hitherto unique local platform for tunable axial gauge fields of Weyl fermions in a three-dimensional solid. This makes the *R*AlGe system particularly promising to explore this physics, since the electronic structure, magnetism and nature of Weyl nodes is expected to depend strongly on the magnetic rare earth ion *R*[12,46,47]. Therefore chemical substitution promises a straightforward route towards an easy low magnetic field control control of both Weyl node type and possible axial gauge fields of Weyl fermions, complementing other control methods such as rotation of a uniform magnetization[48], photo-induction[49,50], or external pressure-induction[51], and thus adding to the catalogue of functional responses that may be useful for applications.

During completion of this work we became aware of another study investigating the anomalous Hall effect in PrAlGe[52].

**Methods**

**Sample synthesis and characterization**

PrAlGe single crystals were obtained by both flux-growth and floating-zone growth techniques[18]. As discussed in Ref. 18, crystals grown by the two different approaches exhibit similar physical properties, with small differences in characteristic transition fields and temperatures arising due to slight variations in stoichiometry. For the present study, energy dispersive x-ray spectra (EDS) analysis shows flux-grown crystals are typically $Pr_{1.0(1)}Al_{1.14(1)}Ge_{0.86(1)}$ with an Al excess and Ge deficiency with respect to the intended 1:1:1 stoichiometry. Floating zone-grown crystals are closer to the 1:1:1 stoichiometry within uncertainty, with typical composition $Pr_{1.08(24)}Al_{0.97(7)}Ge_{0.95(17)}$.

In the present study, the flux-grown crystals display $T_c \sim 16$ K and $B_s(T = 2$ K$) \sim 0.4$ T, while floating-zone crystals display $T_c \sim 15$ K and $B_s(T = 2$ K$) \sim 0.3$ T. Neutron and transport data have been obtained on samples prepared by both methods and are found to be in good general agreement.

**Powder neutron diffraction (PND)**

PND was performed using the DMC instrument at the Paul Scherrer Institute (PSI). The neutron wavelength was 2.457 Å. Diffraction profiles were collected from a 2 g powder sample grown from flux, and loaded into a standard cylindrical vanadium container. The sample temperature was controlled using a standard Orange cryostat with a base sample temperature 1.6 K. No magnetic field was applied. The refinement of the PND profile in Fig. 2b was performed using the FullProf software[53]. In the software, the instrumental contributions to the diffraction peak shape are taken into account by using fixed peak profile and shape parameters determined from standard sample measurements done in the same instrument setup.

**Small-angle neutron scattering (SANS)**

SANS measurements were performed using the SANS-I instrument at PSI. Most of the SANS data presented here were obtained on a 25 mg floating zone grown single crystal. For the experiment the crystal was mounted with a horizontal plane defined by orthogonal [001]-[100] tetragonal axes and [010] vertical. The [001] axis was aligned approximately with the incoming neutron wavevector $k_i$, and loaded into a horizontal field cryomagnet installed at the SANS beamline such that $\mu_0 H \parallel c \parallel k_i$. The base sample $T$ was 1.9 K.

SANS measurements made use of two instrument configurations to measure magnetic scattering over an extended range of momentum transfer. 1) an incident beam with neutron wavelength $\lambda_n = 8$ Å ($\Delta\lambda/\lambda = 10\ \%$) collimated over a distance of 18 m before the sample, with the scattered neutrons detected by a two-dimensional multi-detector placed 18 m behind the sample, and 2) an incident beam with $\lambda_n = 8$ Å ($\Delta\lambda/\lambda = 10$ %) collimated over 8 m before the sample, and scattered neutrons collected by the multi-detector placed 8 m behind the sample. Data measured in each configuration were normalized with respect to the incident beam intensity to cover a $q$-range 0.007 Å$^{-1}$ < q < 0.055 Å$^{-1}$. Exploratory measurements done upon rotating the sample and magnet together (`rocking' measurements) showed no angle-dependence of the scattered magnetic intensity on the detector over a range of ±5°. Therefore, we collected data at fixed sample angle, with a data collection time at each stabilized $T$, $\mu_0H$ typically of 5 to 20 min. To obtain the one-dimensional profiles in Fig. 2c, the scattered intensity at constant magnitude in |$q$| on the two-dimensional multidetector was integrated over 360° around the beam axis (so-called azimuthal averaging). The SANS data reduction and analysis was performed using the GRASP[54] and SASfit[55] softwares, respectively.

**Bulk magnetic and electrical measurements**

Bulk magnetization, electronic, and thermoelectric transport properties have been explored using commercial Quantum Design Magnetic and Physical Property Measurement Systems (MPMS and PPMS). For the resistivity experiments a Hall-bar electrical contact geometry was created. Good electrical contacts (~1 Ω) were established using DuPont 6838 silver paste cured at 500 °C for 10 min and subsequent application of short high voltage pulses. For the thermoelectric transport a home-built insert for the PPMS was used. The temperature gradient in the sample was held at ≈ 3% of the sample temperature and was measured with Cernox thermometers[56], while the voltage was measured using nanovoltmeters. For all measurements magnetic fields were applied along the easy-*c*-axis.

**First-principles calculations**

Electronic structure calculations were performed using the VASP code[57,58], within the GGA-PBE approximation[59] for the exchange-correlation functional and employing the projector augmented-wave method[60,61]. The Hubbard on-site energy U = 4 eV was used to push the f-electrons of Pr away from the Fermi level. In order to perform integration in (1a) over a fine grid of k-points, we construct the maximally localized Wannier functions (MLWF)[62] by means of the Wannier90 code [63-65]. The MLWF are chosen to reproduce the d and f electrons of Pr, as well as sp3 states of Al and Ge, and the upper edge of the frozen window for the band-disentanglement procedure[66] is fixed at 1 eV above the Fermi level. Finally Eq. (1) is evaluated by means of the Wannier interpolation procedure[67], that we implemented in our Wannier19 code [68]. Wannier19 is a new Python code for evaluation of Brillouin-zone integrals of Berry curvature-related properties, that closely follows the postw90.x module of Wannier90 Fortran code, but with important improvements, that significantly speed-up the calculations and improve the result quality. Among the advances are usage of fast Fourier transform, explicit account of symmetries, and a recursive-adaptive refinement procedure, which recursively increases the density of *k*-points sampling around the points which give the major contribution. Technical details of the implementation of Wannier19 will be published elsewhere, and the code is freely available on github. The evaluation of Eq. (2) is much more susceptible to computational inaccuracies than the AHE in Eq. (1), in particular in terms of the BZ sampling at which the AHC is calculated. We find that to get a converged value of $\alpha_{xy}^A/T$, we need a hyperfine grid of 432x432x432 *k*-points with a subsequent recursive refinement of 3% of the *k*-points.

**Data Availability**

All experimental data presented in the figures that support the findings of this study are available at the Zenodo online repository with identifier doi: 10.5281/zenodo.3568739 .

**Code availability**

The Wannier19 code used for the first principle calculations of the anomalous transport is freely available without restriction online at: https://github.com/stepan-tsirkin/wannier19.

**Acknowledgements**   The authors thank B. Fauqué for fruitful discussions. D.D., L.D., and J.C. acknowledge support by the Swiss National Science Foundation (SNF). D.D. acknowledges further support by the Forschungskredit of the University of Zurich, grant no. FK-18-088. L.D. is partially funded by the Swiss Government through Excellence scholarship. J.S.W. acknowledges funding from the SNF Sinergia Network "NanoSkyrmionics" Grant No. CRSII5_171003, and the SNF project 200021_188707. S.S.T. and T.N. acknowledge funding from the European Research Council (ERC) under the European Union's Horizon 2020 research and innovation programme (ERC-StG-Neupert-757867-PARATOP). A.G.G. acknowledges financial support from the ANR under the grant ANR-18-CE30-0001-01 and the European Union's Horizon 2020 research and innovation programme under grant agreement No 829044.

**Competing interests**   The Authors declare no Competing Financial or Non-Financial Interests.

**Author Contributions**   The project was conceived and led by J.S.W., J.C. and T.N. Bulk magnetic and transport measurements were performed and analysed by D.D., L.D., Y.X., J.C. and A.S. Neutron experiments and data analysis were performed by J.S.W., L.K. and J.K. J.K. developed the models for interpreting the SANS data. Theoretical calculations were done by S.S.T., T.N. and A.G.G. Samples were prepared by P.P. and E.P. The paper was prepared by D.D., S.S.T., J.C., T.N. and J.S.W. All authors reviewed the manuscript and agree with the results and conclusions.

**Additional Information**  Supplementary Information accompanies the paper online.

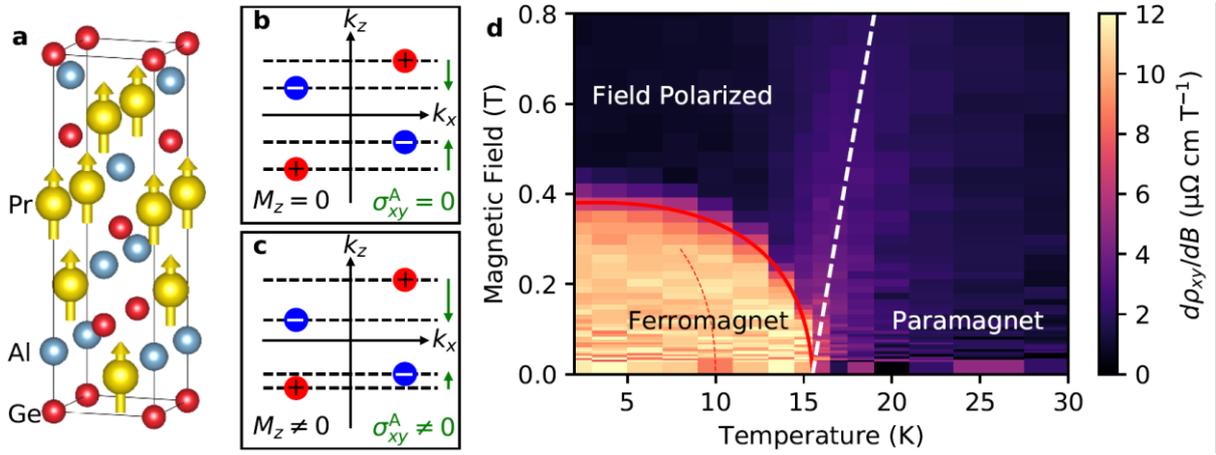

**Figure 1 │ Crystal and magnetic properties of PrAlGe and magnetic response of Weyl nodes. a**, Polar tetragonal PrAlGe crystal structure (space group $I4_1md$, Nr. 109) [15]. Arrows indicate localized magnetic Pr moments aligned with the $c$-axis. Image made using VESTA [16]. **b,c** Schematic of Weyl node locations in momentum space **b** with no magnetization, and **c** in the presence of a spontaneous magnetization $M_z$. The magnetization shifts the Weyl nodes along directions determined by their chirality, contributing to an observable anomalous Hall effect (indicated by green arrows). **d** Phase diagram for PrAlGe in magnetic fields applied along the $c$-axis constructed from the present study. For magnetic fields below the saturation field $B_s \sim 0.4$ T, and temperatures below $T_c \sim 16$ K, a predominantly easy-axis ferromagnetic state is found by neutron scattering. This regime also hosts a large field-derivative of the Hall resistivity found in transport measurements, as shown by the colormap. The thick red solid line guides the eye for the boundary of the ferromagnetic phase. The thin red dashed line indicates a spin-freezing transition identified from magnetization measurements. The white

dashed line denotes a crossover between paramagnetic and field polarized regimes.

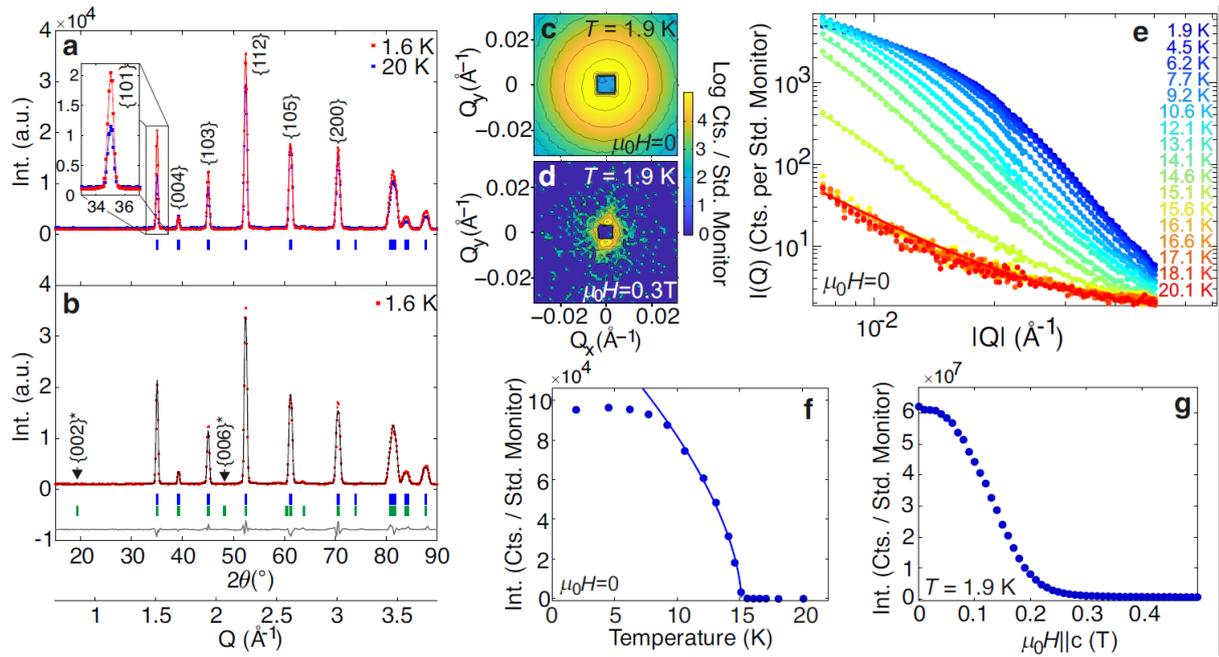

**Figure 2 | Neutron scattering from PrAlGe**. **a** PND profiles obtained at DMC from PrAlGe at base $T = 1.6$ K below $T_c$, and at 20 K above $T_c$. Blue ticks denote the expected positions for nuclear peaks in the $I4_1md$ space group. The inset shows a zoom of the {101} peak. **b** Refinement of the profile measured at $T = 1.6$ K (black line). The dark grey line shows the difference between the refined model profile and the data. The blue ticks denote the expected positions for nuclear peaks, and the green ticks denote possible positions for magnetic peaks according to the magnetic structure model. In both **a** and **b**, positions due to certain scattering planes are labelled. In **b**, * denotes scattering planes only allowed according to magnetic symmetry, the absence of intensity at the labelled peaks implies directly that the magnetic moments are aligned along *c*. The good profile refinement using fixed diffraction peak profile and shape parameters (see Methods) indicates the main scattering peak profiles

are close to resolution-limited. SANS patterns obtained from PrAlGe at 1.9 K in **c** $\mu_0H = 0$, and **d** $\mu_0H = 0.3$ T, each after ZFC. A square beamstop is placed at the center of the reciprocal space $q = 0$ to protect the detector from saturation due to the unscattered beam. The [001] axis is out of the page, and the [100] axis horizontal. Contour lines are plotted in steps of 0.5 on the logarithmic intensity scale, with the highest intensity contour drawn for Log Intensity = 4.5 counts per standard monitor (Cts. / per Std. Monitor). **e** The log-log plot of the azimuthal-averaged SANS intensity versus $|q|$ as a function of $T$ in $\mu_0H = 0$. The solid lines are fits of the data according to the model described in the text and the Supplementary Note 3. **f** The total magnetic scattering contribution to the SANS profiles shown in **e** as a function of $T$ in $\mu_0H = 0$. The solid line is a power law fit. **g** A $\mu_0H$ -increasing scan of the magnetic SANS intensity at 1.9 K. In panels **a**, **b**, **e**, **f**, **g**, error bars indicating the statistical error are smaller than the symbol size.

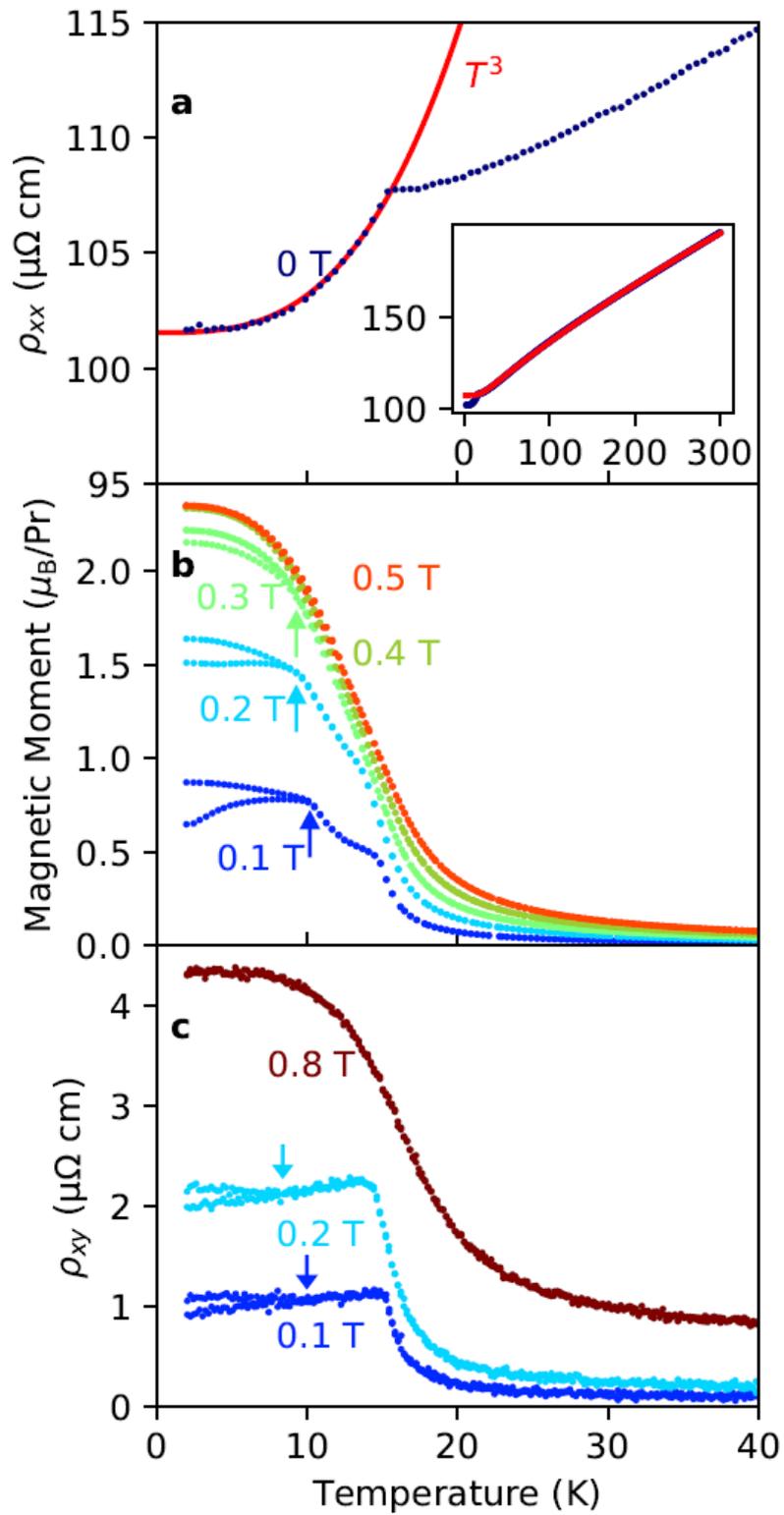

**Figure 3 | Thermodynamic and transport measurements of PrAlGe. a** Low $T$ resistivity showing a clear cusp at $T_c$ due to magnetic order. The red line shows a $T^3$ scaling of the

resistivity up to $T_c$. The inset show the $T$-dependent resistivity over a wider $T$ range and a fit to the Bloch-Grüneisen formula minus a $T^3$ correction term. **b** The magnetic moment as a function of temperature for magnetic fields as indicated. Saturation magnetization is reached at 0.35 T. Each curve was measured zero-field-cooled (ZFC) and field-cooled (FC). Below the saturation field, the two curves bifurcate at $T_g$ ~10 K (indicated by arrows), and the ZFC branch has a lower moment than the FC branch. **c** Hall resistivity as a function of $T$ and magnetic field, showing clear resemblance to the magnetization in panel **b**.

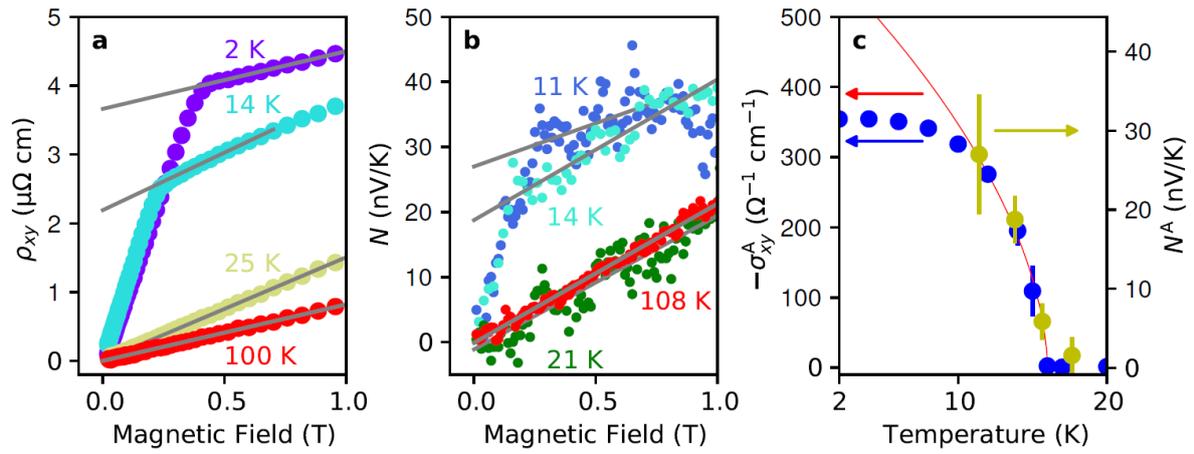

**Figure 4 | Anomalous transport in PrAlGe. a** The Hall resistivity $\rho_{xy}(\mu_0 H)$ for various $T$. Below $T_c$, a kink at magnetic saturation is observed that gets more pronounced at lower $T$. Grey lines indicate extrapolations to zero magnetic field. **b** Nernst isotherms for temperatures as indicated, resembling the behaviour of the Hall isotherms. **c** The $T$-dependence of the anomalous Hall conductivity $\sigma_{xy}^A$ and Nernst effect $N^A$ extrapolated to zero magnetic field from data measured above magnetic saturation. Error bars are obtained by varying the fitting window to account for slight curvature. The red line is a fit of $\sigma_{xy}^A$ to a power law. Where visible, error bars indicate the standard error.

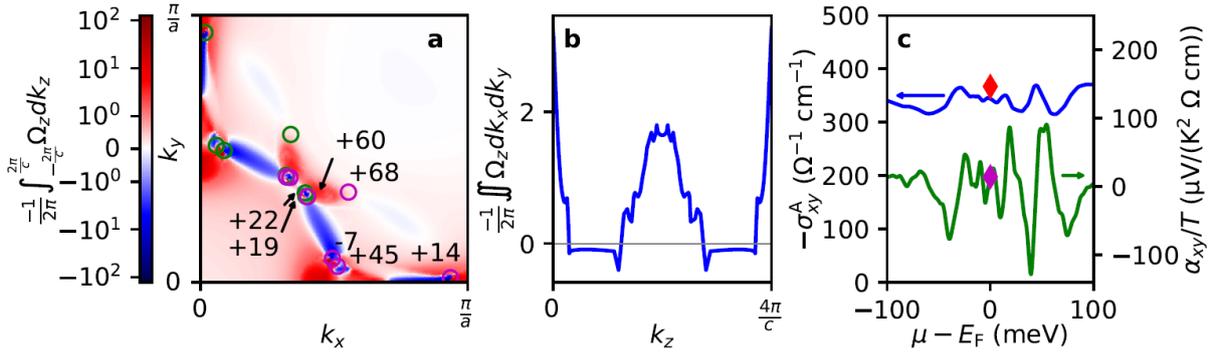

**Figure 5 │ Calculation of anomalous Hall conductivity in PrAlGe. a** The negative Berry curvature integrated along $k_z$ as a function of $k_x$, $k_y$. Due to the $C_4$ rotational symmetry of PrAlGe, we show only a quarter of the $k_x$-$k_y$ plane. The circles indicate the projections of the Weyl nodes. Green indicates positive chirality and magenta negative chirality. The numbers correspond to the distance to the Fermi level in meV. Only one circle of each symmetric pair is indicated with a number. **b** The negative Berry curvature (summed over states below the Fermi level) integrated in the $k_x$-$k_y$ plane in the range $-\pi/a \leq k_x, k_y \leq \pi/a$, as a function of $k_z$. **c** Dependence of the calculated $\sigma_{xy}^A$ at $T = 0$ (blue line, left axis) and $\alpha_{xy}/T$ at $T = 14$ K (green line, right axis) on the position of the chemical potential relative to the calculated Fermi level. The diamonds denote the experimentally determined values of $\sigma_{xy}^A$ at 2 K (red diamond) and $\alpha_{xy}/T$ at 14 K (magenta diamond).